# A Framework for Automated Pop-song Melody Generation with Piano Accompaniment Arrangement

Ziyu Wang[1,2], Gus Xia[1]

[1]New York University Shanghai, [2]Fudan University

{ziyu.wang, gxia}@nyu.edu

**Abstract:** We contribute a pop-song automation framework for lead melody generation and accompaniment arrangement. The framework reflects the major procedures of human music composition, generating both lead melody and piano accompaniment by a unified strategy. Specifically, we take chord progression as an input and propose three models to generate a structured melody with piano accompaniment textures. First, the harmony alternation model transforms a raw input chord progression to an altered one to better fit the specified music style. Second, the melody generation model generates the lead melody and other voices (melody lines) of the accompaniment using seasonal ARMA (Autoregressive Moving Average) processes. Third, the melody integration model integrates melody lines (voices) together as the final piano accompaniment. We evaluate the proposed framework using subjective listening tests. Experimental results show that the generated melodies are rated significantly higher than the ones generated by bi-directional LSTM, and our accompaniment arrangement result is comparable with a state-of-the-art commercial software, Band in a Box.

## 1 Introduction

In recent years, great progress has been made in music automation with the development of machine learning. Various generative models have been able to generate interesting music segments. To name a few, [6, 8, 10] for melody generation and [20] for accompaniment arrangement. However, most models merely focus on specific modules of music generation and rarely consider how to connect or unify the modules in order to generate a complete piece of music. To be specific, we see three severe problems. First, melody generation and polyphonic accompaniment arrangement are mostly treated two separate tasks. Consequently, melody generation models cannot be applied to generate voices in the polyphonic accompaniment directly as composers usually do. Second, end-to-end sequence-generation models lack the representation and design of phrasing structure, resulting in "noodling around" music. Last but not least, a given chord progression is regarded a rigid input of music generation systems, instead of a soft constraint that is flexible to be altered by composers to interact with different music styles.

To solve the above three problems, we contribute a pop-song automation framework for lead melody generation and accompaniment arrangement (as shown in **Fig. 1**). The framework follows the major procedures of human music composition and generates melody and accompaniment in a unified strategy. A popular song usually consists of three parts, namely a *chord progression*, a *lead melody*, and an *accompaniment* (represented by the three corresponding dotted rectangular areas). The framework uses three models to execute the generation process, namely *harmony alternation model*, *melody generation model* and *melody integration model* (represented by corresponding colored arrows).

We assume a minimum input of a raw (original) chord progression for the whole framework, which can be either manually defined or automatically generated using harmonization algorithms.

In the first step, the harmony alternation model transforms the raw, original progression into a concrete, decorated one to best fit a certain music style. The underlying idea is that any initial progression should only be a "soft" restriction of a piece and adaptable to different music context. For example, a major triad could be modified as an "add6" chord for Chinese music or as major 11$^{th}$ chord for jazz music. The second step is the most important one, in which the melody generation model considers the accompaniment a set of melodies (monophonic melody lines, e.g. secondary melody, arpeggios etc.), performs a hierarchical contour decomposition to each melody, and generates melodies in parallel using seasonal ARMA (Autoregressive Moving Average) processes [17]. The core feature of this step is that the model can create lead melody in exactly the same way, and hence unifies melody generation and accompaniment arrangement problems. Finally, the melody integration model combines melodies into parts (e.g. left-hand part and right-hand part) and adds column chords to embellish the accompaniment.

The rest of the paper is structured as follows. The next section presents the related work. We present the methodology in Section 3, show the evaluation in Section 4, and conclude the work in Section 5.

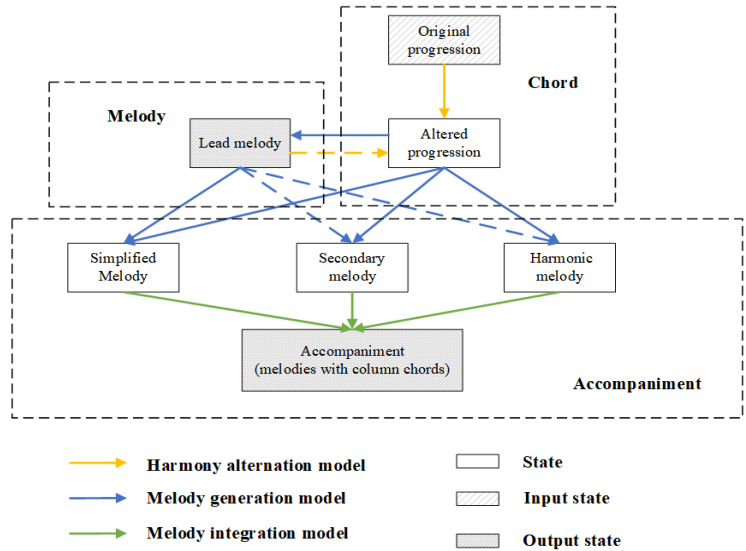

**Fig. 1 A system diagram of the melody generation and accompaniment arrangement framework. The dotted arrows are to be implemented.**

## 2  Related Work

We review three realms of related work, namely chord progression generation, melody generation and accompaniment arrangement.

A chord can be represented as nominal [11], continuous [13], or structural [4] variables. A nominal representation builds a straightforward one-to-one mapping between pitches and chord symbols. Such simple representation has been used in various tasks, such as chord extraction [3, 11], melody harmonization [15, 18], and automated composition [9]. To reveal chord distance, chord embedding representation (say, in a continuous psychoacoustic space) has been proposed to address jazz harmonization [13]. The work by Cambouropoulos et al. [4, 12] further used a hierarchical chord structure to reveal chord similarity from an analytical perspective. Based on the idea of [4], our model performs structural chord alternation on original progressions in order to better match different music styles. To generate a chord progression, the common approaches are directed probabilistic graphical models [15, 18] and tree structured models [19]. The target of these models is to find the optimal chord progression that is arranged in a most logical way and agrees with the input melody most. In the context of automatic composition and accompaniment arrangement, this is the first study to consider the alternation of chords.

Current melody generation methods can be categorized into two types: traditional Markovian approaches [9, 14] and modern deep generative models [1, 8, 10, 16]. According to [7], both approaches are not able to generate truly creative music. The former could interact with human inputs, but requires too much constraint and can hardly capture long-term dependencies. The latter, on the other hand, have long-term memories but still largely depend on training data and cannot yet interact with user input. In our framework, we use weaker constraint for melody generation, providing an insight on the connection of Markovian and deep learning approaches in the future.

For melody generation and accompaniment arrangement, the framework of XiaoIce band [20] is very relevant to our work. It used two end-to-end models for lead melody generation and accompaniment arrangement respectively. The first model used a melody-rhythm-cross-generation method to improve its rhythmic structure, while the second model use multi-task joint generation network to ensure the harmony among different tracks. Compared to XiaoIce band, we used a unified strategy to generate both lead melody and voices in the accompaniment. Moreover, we are more focused on revealing the music composition procedures in automated generation including chord progression re-arrangement and music structural design.

## 3 Methodology

In this section, we present the design of our framework in detail. As shown in **Fig. 1**, given an original chord progression, the entire generation process contains three steps, each associated with a tailored model. We present the harmony alternation model in Section 3.1, present the melody generation model, which is the core part of the framework in Section 3.2, and present the melody integration model in Section 3.3.

### 3.1 Harmony Alternation Model

In most music automation systems, chord progressions are taken as rigid inputs without any changes. However, according to music theory, a chord progression is a guideline rather than a fixed solution. This is analogous to a recommended GPS route, which has to be adjusted based on various traffic situations. For example, a progression of [C, Am, F, G] can be altered into [Cmaj7, Am7, Fmaj7, G7] for jazz music. For pop songs, [Cadd2, Am7, Fmaj7(add 9), Gsus4] is more likely to appear.

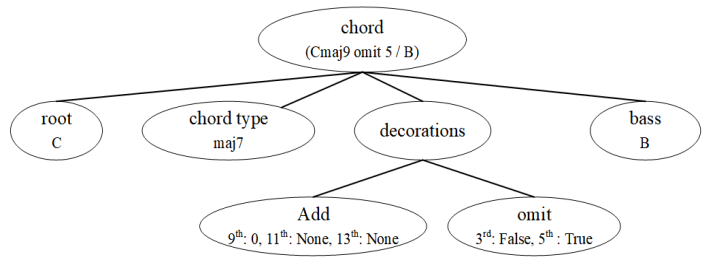

**Fig. 2 An example of the proposed chord representation.**

### 3.1.1 Chord Representation

A chord may contain many notes, and it is generally considered that only a subset (usually the first three/four notes) of the chord decides its basic type and function, whereas the other notes make the chord more complicated and characterized. Inspired by this observation, we represent a chord by four parts: *root*, *chord type*, *decorations* and *bass*.

**Fig. 2** shows an example of the four-part chord representation. *Root* is the lowest note in a chord, which is denoted by one of the 12 pitch classes. *Chord type* is a simplified version when a chord is reduced to triads or seventh chords (which decides the basic types and functions). *Decorations* consists of two sub-parts: *add* and *omit*. The former records whether 9th, 11th, and 13th are in the chord and the interval to the default degrees (major 9th, perfect 11th, and major 13th respectively) in semitones. The latter records whether the 3rd and 5th degree note is omitted in a chord.

### 3.1.2 Chord Decoration Operations

We currently define two chord-decoration operations: `add()` and `omit()`. The two operations add or reduce the corresponding note indicated in the brackets. For example, when `add((11th,0))` is applied to a major triad, the chord becomes an add4 chord, whereas if `add((11th,1))` is applied to a major seventh chord, the chord becomes a 11th (#11, omit 9) chord. In the same sense, when `omit(1st)` operation is applied to Em7, it becomes a G; if `omit(3rd)` is applied to Em7, it becomes Em7(omit3); and if `omit(9)` is applied to it, nothing happens. These operations keep track of how much the chord has changed. For example, a modification of the root note is considered a large change, while 11th and 13th of a chord have smaller impact on the chord function.

### 3.1.3 Decorate Chord Progression

Based on the chord representation and the definition of decoration operation, an altered chord progression can be obtained by the original progression with a sequence of decoration operations. To model their relationship, an HMM is trained based on 890 songs in McGill Billboard dataset [3].

Formally, let $[d_1, d_2, ..., d_n]$ be the hidden (decorated) state where each $d_i$ is an (`add`, `omit`) tuple we ever observed from the dataset. Let $[c_1, c_2, ..., c_n]$ denote the original chord sequence, where each $c_i$ contains three parts: the root $c_i^r$, chord type $c_i^{ct}$, and chord duration $c_i^t$. Hence, the transition probability is $p(d_i|d_{i-1})$, while the emission probability is defined as the product of three terms: chord type emission $p(c_i^{ct}|d_i)$, duration emission $p(c_i^t|d_i)$, and local chord connection emission $p(c_i^r - c_{i-1}^r, c_{i+1}^r - c_i^r|d_i)$. We learn these probabilities directly from data, perform Viterbi algorithm to decode the top $N$ decorated chord progressions and choose the most suitable progression by a self-defined optimization function.

## 3.2 Melody Generation Model

Melody generation model is the core part of the framework. By melody, we do not only mean the lead melody, but rather, in a general sense, every discernable monophonic component in the composition. To be specific, we decompose a pop song into four types of melody lines (or melody), namely *lead melody*, *simplified melody*, *secondary melody* and *harmonic melody*. Lead melody is the human voice. Simplified melody supports the lead melody, which is a variation of lead melody with less notes and less complicated rhythmic pattern. Secondary melody serves as a parallel theme, an independent melody. Harmonic melody reflects the chord progression, which is usually specific patterns of broken chords, including arpeggio, walking bass, Alberti bass, etc. **Fig. 3** shows an example, in which the upper part is the original composition and the lower part is the decomposed melodies. In this section, we discuss how to use a unified model to generate the four types of melodies.

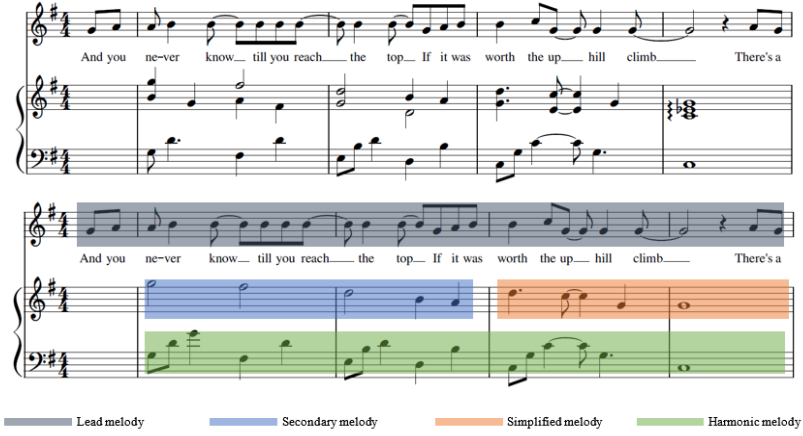

**Fig. 3 A comparison between the original accompaniment and its decomposed melodies.**

### 3.2.1 Melody Representation

We denote a melody $\{m_i\}_{i=1}^n$, as a time series of length $n$, in which each item can be one of the 128 MIDI pitches and two extra states: *silence* and *sustain*. Each timestamp corresponds to a 16$^{th}$ note. We compose melody into two parts:

$$m = c + \epsilon, \tag{1}$$

where $c = \{c_i\}_{i=1}^n$ is the *contour*, $\epsilon = \{\epsilon_i\}_{i=1}^n$ is the (quantization) *error*. Both are real-value vectors. For example, $n = 64$ if the time signature of a 4-bar melodic phrase is 4/4.

This simple form reflects two different procedures in composing. The contour term $c$ is a continuous, preliminary blueprint of the melody, which is analogous to a composer's inspirations and suitable to model phrase-level structure. The error term $\epsilon$ transforms approximate contours into accurate MIDI pitches, which is analogous to using domain knowledge to realize inspirations into actual notes. $\{\epsilon_i\}_{i=1}^n$ are not i.i.d but correlated, which will be fully explained in Section 3.2.3.

### 3.2.2 Contour-inspiration Model

The contour-inspiration model divides $c$ into components called *layered signals* denoted by $s^{(k)} = \{s_i^{(k)}\}_{i=1}^n$:

$$c = s^{(0)} + \sum_{k=1}^p s^{(k)}, \tag{2}$$

where $i$ is the index of element within a layered signal, and $k$ is the index of layered signals. $s^{(0)}$ is a deterministic trend, and $s^{(k)}, k = 1, 2, \ldots, p$ are stochastic processes. These stochastic processes describe the shape of the melody contour in various periods. Particularly, the layered signal $s^{(k)}$ has a sample rate $2^k$ and captures only the contour information not obtained in the previous layers with lower sample rates. In this way, a melody contour is decomposed into different seasonal components.

For a given melodic phrase $m$, the decomposition procedure is as follows: $s^{(0)}$ is a deterministic trend, and we keep $s_1^{(0)} = m_1$. For other layers, we first introduce intermediate variables (layers) $x^{(k)}, k = 1, \ldots, p$, which is melody $m$ in sample rate $2^k$. Then, $s^{(k)}$ is defined as $s^{(1)} = x^{(1)} - s^{(0)}$ when $k = 1$ and $s^{(k)} = x^{(k)} - x^{(k-1)}$ when $k = 2, \ldots p$. It is apparent that $m$ is the summation of the layered signals $s^{(0)}, s^{(1)}, \ldots, x^{(p)}$. Moreover, for a single layer, when the sampled timestamps overlap with the previous layer, the values will naturally be zeros (see **Fig. 4** ). This decomposing idea is inspired by the observation that phrase-level melodic structure is usually symmetrical and exists in different time scales. In order to generate a melody, we model the melody contour via simulating these layers with different stochastic models, e.g. deterministic process, ARMA process, white noise process, etc.

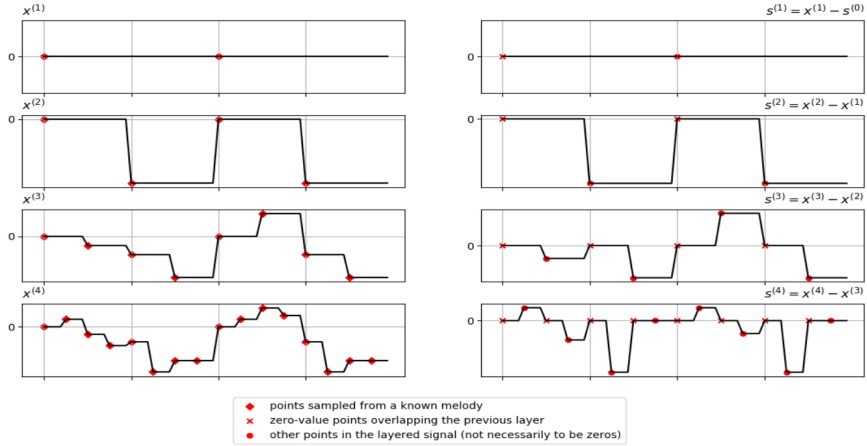

**Fig. 4 Demonstration of melody contour under different sample rate (left) and layered signals (right).**

### 3.2.3 Error-expertise Model

The contour-inspiration model $c$ generates a continuous melody contour, while error expertise model $\epsilon$ performs quantization based on domain knowledge. In theory, $\epsilon$ is a correlated multivariate Gaussian distribution (weighted by chord context). In practice, the distribution is weighted by chord context and modified by rules in the following to ways.

First, the model quantizes the contour (floating points) into MIDI pitches (integers) under the context of chord progression. Specifically, an exact MIDI pitch is selected under a Gaussian distribution (centered at the contour float) weighted by $p(pitch|chord)$ learned from data.

Second, the model adjusts the rhythm of the melody contour. Rather than assigning a rhythmic pattern, we derive rhythms from the melody contour. Generally, when two adjacent contour values are closer than a threshold, we merge the two notes, assigning a sustain state to the latter one.

From section 3.2.4 to section 3.2.7, we discuss how to apply melody generation model (2) to the four types of melodies introduced in the beginning of section 3.2.

### 3.2.4 Lead Melody Generation

In contour-inspiration model (2), $s^{(0)}$ is a constant and $s^{(1)}, \dots, s^{(6)}$ are modeled by seasonal ARMA $(1,1) \times (1,1)_s$ (seasonal Autoregressive Moving Average) processes [17] with different parameters. The hyperparameters (i.e., the order of the model) are set based on the observation of the ACF (autocorrelation function) and PACF (partial autocorrelation function) of the layered signals.

In error-expertise model, melody contour $c$ is quantized to discrete MIDI pitches under a weighted Gaussian distribution (see Section 3.2.3). Before that, we use rules to make sure adjacent notes with similar contour values are quantized to the same pitch. Particularly, we adopt a threshold $\eta$; if $|c_i - c_{i-1}| < \eta$, $m_i = sustain$; otherwise, we select the note according to the distribution given above.

### 3.2.5 Secondary Melody Generation

The model is exactly the same as lead melody generation. In modeling seasonal ARMA process, we set the parameters within a low range since secondary melody is usually less complicated than the lead melody.

### 3.2.6 Harmonic Melody Generation

In contour-inspiration model, in order to represent the accompaniment texture, we learn a deterministic trend $s^{(0)}$ from pattern samples. We assume $s^{(0)} = bass + pattern$, in which $bass$ is the bass of the ongoing chord and $pattern$ a particular way to arrange notes into sequence. $bass$ is extracted from the input chord progression, and $pattern$ is estimated by the sample means. As for the other layered signals, they are modeled as white noises to improve randomness and enhance the sense of improvisation. Error-expertise model is the same as used in Section 3.2.4.

### 3.2.7 Simplified Melody Generation

In simplified melody generation, contour-inspiration model captures the shape of the melody to be simplified and the error-expertise model executes the simplification. Specifically, in contour-inspiration model, the deterministic trend $s^{(0)}$ is identical to the lead melody and other $s^{(k)}$ are set to be zero. In error-expertise model, it makes delete-note and alter-onset decisions on each $c_i$ according to various properties, such as passing note, trill, downbeat, etc. We grade each note their importance and delete the relatively unimportant notes. Also, some note onsets are moved to the downbeat if the note supposed to be at that beat position is deleted. In short, decorative notes as well as outliers are likely to be deleted, whereas critical notes that shape the contour of the melody are maintained.

## 3.3 Melody Integration Model

Melody integration model acts as the final step in our system. For now, this model is only in its preliminary phase, which consists of a set of rule-based algorithms.

First, it combines secondary melody, simplified melody and harmonic melody together as the accompaniment. In our current settings, harmonic melody serves as the left-hand part. As for the right-hand part, a rule-based method is designed to combine secondary melody and simplified melody. We use function to analyze the smoothness of lead melody per bar. If the smoothness exceeds a threshold, secondary melody is selected. Otherwise, simplified melody is selected to support the lead melody. Second, column chords are randomly added to the accompaniment based on a manually defined distribution. Notes with higher pitch or a relatively strong metrical strength are likely to be appended with column chord.

# 4 Experimental Results

We evaluate the performance of melody generation and the whole system through listening experiments. We created a survey to evaluate melody generation model and the accompaniment arrangement system. We compared the former with a bi-directional LSTM model [6] (the representative deep generative model for music generation). We compared the latter with Band in a box (short for BIAB, the state-of-the-art commercial software for accompaniment generation). We showed the paired music demos in a random order to each experimenter without directly revealing the condition. Each demo is about 30 seconds long. After each demo, they are asked to rate the overall musicality, interactivity (between melody and progression/accompaniment and melody), structural organization. For all three criteria, we used a 5-point continuous scale from 1 (very low) to 5 (very high).

We collected 47 and 41 valid samples for melody comparison and accompaniment comparison. To validate the significance of differences, we conducted paired t-test. **Fig. 5** shows that our model is significantly better than LSTM model (with p-values $< 0.005$) for melody generation and marginally better than BIAB (p-values $> 0.5$ ) for accompaniment arrangement.

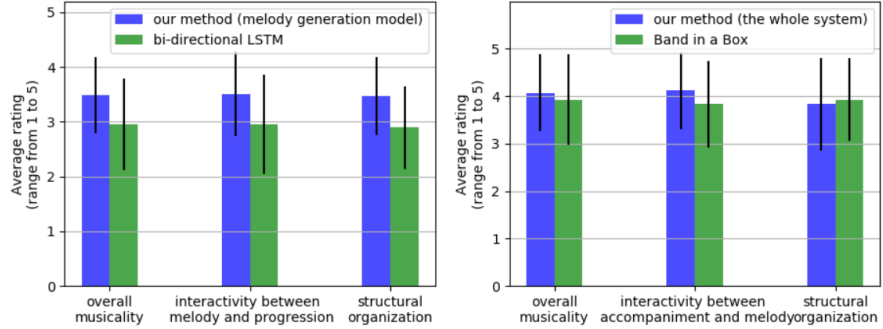

**Fig. 5 A comparison between our framework with LSTM for melody generation (left) and BIAB for arrangement (right).**

We provide demos for each step in our system as well as the overall generation. Demos are available at demo-album: https://soundcloud.com/wang-michael-452158298/sets/auto-compose.

## 5  Conclusion and Future Work

We have created an automated composition framework. Firstly, we improve the existing chord model to enhance the inner-relationship among chords. Secondly, we decompose the whole composition into a set of melodies and regard each melody generation as a two-step procedure by dividing melody model into two separate sub-models. Last but not least, we present a method to integrate melodies into one whole composition.

An ideal framework should be able to understand both concrete and abstract music content and interact with people at different levels of abstraction. We see our framework a first attempt towards this goal. In future, we plan to 1) conduct more analysis on the connection between melody contour signals and error-expertise model, 2) explore more effective representation of music structure, and 3) design better methods for melody integration.